\documentclass[pra,twocolumn,tightenlines,showpacs,nofootinbib]{revtex4}
\usepackage{bm,dcolumn,amsmath,graphicx}

\newcommand{\eref}[1]{Eq.~(\ref{#1})}
\newcommand{\tref}[1]{Table~\ref{#1}}

\begin{document}

\title{Excitation of the isomeric $^{229m}$Th nuclear state
via an electronic bridge process in $^{229}$Th$^+$}

\author{S. G. Porsev$^{1,2}$}
\author{V. V. Flambaum$^1$}
\author{E. Peik$^3$}
\author{Chr. Tamm$^3$}
\affiliation{$^1$ School of Physics, University of New South Wales,
Sydney, NSW 2052, Australia}
\affiliation{$^2$ Petersburg Nuclear Physics Institute, Gatchina,
Leningrad district, 188300, Russia}
\affiliation{$^3$ Physikalisch-Technische Bundesanstalt,
Bundesallee 100, 38116 Braunschweig, Germany}

\date{ \today }
\pacs{31.15.A-, 23.20.Lv, 27.90.+b}

\begin{abstract}
We consider the excitation of the nuclear transition $^{229g}$Th -- $^{229m}$Th near 7.6 eV in singly ionized thorium via an electronic bridge process. The process relies on the excitation of the electron shell by two laser photons whose sum frequency is equal to the nuclear transition frequency. This scheme allows to determine the nuclear transition frequency with high accuracy. Based on calculations of the electronic level structure of Th$^+$ which combine the configuration-interaction method and many-body perturbation theory, we estimate that a nuclear excitation rate in the range of 10 s$^{-1}$ can be obtained using conventional laser sources.
\end{abstract}

\maketitle

The nuclear transition between the low-energy isomeric state and the
ground state in the $^{229}$Th nucleus is very interesting
due to the possibility to build a very precise nuclear clock and
its high sensitivity to a hypothetical temporal variation of the fundamental
constants~\cite{PeiTam03,Fla06}.
The value of the transition frequency is known from $\gamma$-spectroscopy \cite{KroRei76,ReiHel90,BecBecBei07} with an uncertainty that is many orders of magnitude higher than the presumed linewidth. The presently most precise value of the transition energy is $7.6(5)$~eV \cite{BecBecBei07}, placing the transition in the vacuum-ultraviolet spectral range.
Further investigations are therefore needed in order to allow a direct optical excitation and detection of the line and to gradually increase the spectroscopic resolution to the level adequate for an optical clock  \cite{PeiTam03}. Since the energy of this nuclear transition is in the same range as resonances of the outer-shell electrons in thorium ions and chemical compounds of Th, the electronic environment may have a significant influence on the nuclear transition rate \cite{Tka03}. In this work we suggest a new experiment to study laser excitation of the nuclear
$^{229g}$Th -- $^{229m}$Th transition, making use of the dense electronic level structure of Th$^+$.

In a recent article~\cite{PorFla10ThII}
we have studied the  effect of atomic electrons on the nuclear transition
from the isomeric $^{229m}$Th state to the ground $^{229g}$Th state
in $^{229}$Th$^+$ due to an electronic bridge (EB) process.
Here we consider the process
of the nuclear $g \rightarrow m$ {\em excitation} via an EB process.
Atomic units ($\hbar = |e| = m_e = 1$) are used unless noted otherwise.

\begin{figure}
 \includegraphics[scale=0.5]{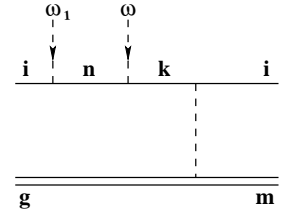}
 \caption{Feynman diagram of the considered two-photon electronic bridge process. The single and double solid lines relate to the electronic and the nuclear states, respectively. The dashed lines are the photon lines.}
 \label{Fig:EB1}
 \end{figure}

{\bf Experimental scheme}.
The EB process considered in this work can be represented by the
Feynman diagram in Fig.~\ref{Fig:EB1}. In the following we assume
a resonant character of this EB process. For this reason we take into account
only one Feynman diagram given by Fig.~\ref{Fig:EB1} which mainly contributes to the
probability of the process. Other diagrams which can be obtained from this one
by permutations of the photon lines will be neglected.
We suppose that the initial state $i$ is the electronic ground state and the initial and final
electronic states are the same. Further we assume a two-photon excitation of
high-lying states $k$. For the first excitation at $\omega_1$, it seems convenient to
use a continuous laser tuned to the transition from the Th$^+$ ground state
$(6d^2 7s)\, J=3/2$ to the $(6d7s7p)\, J=5/2$ state at 24874 cm$^{-1}$ \cite{lac-database}.
This electric dipole (E1) transition is known as the strongest emission line in the Th$^+$
spectrum and it can be excited efficiently in collisionally cooled ion clouds in a radiofrequency
trap \cite{Zimmermann}. As a result, only the $(6d7s7p)\, J=5/2$ state needs to be accounted
for in the sum over intermediate states $n$ in Fig. 1. We also note that the monitoring of
the fluorescence resulting from excitation of the $(6d^2 7s)\, J=3/2 \rightarrow (6d7s7p)\, J=5/2$
transition allows us to detect the excitation of $^{229}$Th$^+$ ions from the nuclear ground state
to the isomeric state, making use of the differences between the hyperfine structure characteristics
of electronic transitions in $^{229m}$Th$^+$ and in $^{229g}$Th$^+$ \cite{PeiTam03}.

The second excitation step with frequency  $\omega$ gives rise to the excitation
of higher-lying states $k$. Their decay to the ground state is accompanied by the nuclear
$g \rightarrow m$ transition and leads to the appearance of isomeric Th$^+$ ions in the electronic
ground state. For this excitation scheme, energy conservation implies that
$\omega_1 + \omega = \omega_N$,
where $\omega_1$ and $\omega$ are the frequencies of the incident photons,
and $\omega_N = E_m - E_g$ is the nuclear transition frequency as determined by
the difference between the isomeric nuclear energy $E_m$ and the ground-state energy $E_g$.

This excitation scheme offers the prospect to determine the nuclear transition
frequency $\omega_N$ with the accuracy afforded by high-resolution laser spectroscopy.
In an experiment based on this scheme, one would use a widely tunable laser source to produce
narrow-bandwidth radiation with variable frequency $\omega$. If the laser frequency is scanned
over the nuclear resonance,
 the probability of excitation
to the isomeric state exhibits a resonance peak. In an ion-trap experiment with collisionally cooled
$^{229}$Th$^+$ ions, the width of the resonance is determined essentially by Doppler broadening
and by the combined linewidth of the employed laser sources. If one assumes saturated excitation
to the state $n$ and a fixed detuning of $\omega$  relative to the electronic transitions
$n \rightarrow k$, the probability of the EB process shown in Fig. 1 is proportional to the spectral
intensity of the laser field at $\omega$.

Based on the result of Ref. \cite{BecBecBei07}
we assume that the most interesting
range of the electronic excitations is around 7.6 eV $\approx$ 61300 cm$^{-1}$.
Following this assumption, we expect that in the sum over the intermediate
states $k$ the atomic energy levels lying between
$60\,000$ and $64\,000$ cm$^{-1}$ will give the predominant
contribution to the probability of the EB process. Unfortunately,
these energy levels are not yet identified experimentally. Therefore
all the following results are based on
{\it ab initio} calculations.
As we noted in~\cite{PorFla10ThII}, the achieved accuracy of calculations
of the high-lying states of Th$^+$ is at the level of several percent. This is not sufficient
to reliably predict
the resonance enhancement occuring in the EB process.
Hence, the
experimental identification of the energy levels should be considered as
the next step towards realizing the method considered here.

\begin{figure}
 \includegraphics[scale=0.5]{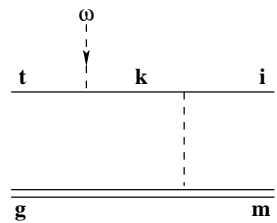}
 \caption{Simplified Feynman diagram of the considered electronic bridge process. Here it is assumed that the first excitation step leads to efficient population of the state $(6d7s7p)\, J=5/2$ at 24874 cm$^{-1}$ which is denoted as $t$.}
 \label{Fig:EB2}
 \end{figure}

Because at present very accurate calculations are not needed we will make
one more assumption simplifying the calculations. We assume that
only one intermediate state $n$, the state at 24874 cm$^{-1}$, contributes to the
probability of the EB process and that 100\% population of this
state can be achieved. As a result, the frequency $\omega_1$ is assumed to be fixed
at 24874 cm$^{-1}$ and
the process which we discuss in the following can be described
by the diagram represented by Fig.~\ref{Fig:EB2}, where
the state at 24874 cm$^{-1}$ is denoted as $t$ and is
considered as the initial state.

{\bf Calculation}. Figure~\ref{Fig:EB2} shows an EB process that relies on the absorption of an incident photon.
As we mentioned in Ref.~\cite{PorFla10ThIV} an EB process of this type can be effectively
treated as a ``generalized'' electric dipole transition from the initial
to the final state.
If the incident radiation with spectral intensity $I_\omega$ is isotropic
 and unpolarized,
the relation between the probabilities $W_{ab}$ of a spontaneous
transition $a \rightarrow b$ and $W^{\rm in}_{ba}$ of the corresponding induced transition $b \rightarrow a$
is given by~\cite{Sob79}
\begin{equation}
W^{\rm in}_{ba} = W_{ab} \,  \frac{4 \pi^3 c^2}{\omega^3} I_\omega .
\label{Wa}
\end{equation}
The spontaneous EB process can be formally described by the mirror-image of
Fig.~\ref{Fig:EB2} with an outgoing photon arrow.
The general formalism was discussed in detail in two recent
articles~\cite{PorFla10ThIV,PorFla10ThII}.
Assuming the resonance character of the EB process
the expression for $\Gamma_{\rm EB}$
can be written as~\cite{PorFla10ThII}:
\begin{eqnarray}
\Gamma_{\rm EB} \approx \frac{4}{9}\left( \frac{\omega}{c}\right)^3
\frac{|\langle I_g||\mathcal{M}_1||I_m \rangle|^2}{(2I_m+1)(2J_t+1)}\, G_2 ,
\label{GamK}
\end{eqnarray}
where $\mathcal{M}_1$ is the magnetic dipole nuclear moment and
$|I_g \rangle = 5/2^+$ [633]  and $|I_m \rangle = 3/2^+$ [631] are the ground and the isomeric nuclear state, respectively,
given in their Nilsson classification.
$J_i$ is the electron total angular momentum of the initial state and
 $\omega = \omega_N - \omega_1$ is the frequency of an absorbed photon.
The explicit expression for the coefficient $G_2$ was derived in~\cite{PorFla10ThIV}
and reads as
\begin{eqnarray}
G_2 &=& \sum_{J_k} \frac{1}{2J_k+1}  \nonumber \\
&\times& \left\vert \sum_{\gamma_s}
\frac{\langle \gamma_i J_i ||\mathcal{T}_1|| \gamma_s J_k \rangle
      \langle \gamma_s J_k ||D|| \gamma_t J_t \rangle}
     {\omega_{si}-\omega_N + i\Gamma_s/2}\right\vert^2\!,
\label{G2}
\end{eqnarray}
where $\omega_{si} \equiv \varepsilon_s - \varepsilon_i$,
and $\varepsilon_l$ denotes the energy of level $l$,
$\mathcal{T}_1$ is the electronic magnetic-dipole hyperfine coupling operator,
$D$ is the electric dipole moment operator,
and $\gamma_s$ encapsulates all other electronic quantum numbers.
The explicit expressions for the matrix elements of the operators $\mathcal{T}_1$
and $D$ are given in Ref.~\cite{PorFla10ThIV}.
The possible values of the total angular momentum $J_k$ are determined by
the selection rules of the operators $\mathcal{T}_1$ and $D$. In particular,
in our case $J_i = 3/2$ and $J_t = 5/2$ and, correspondingly, $J_k$ can be equal to
3/2 and 5/2.
%

In the following we will use the dimensionless quantity $\beta_{M1}$ introduced in
Ref.~\cite{PorFla10ThIV} and defined as the ratio of the probability of the spontaneous
EB process $\Gamma_{\rm EB}$ to the probability of the spontaneous $M1$ radiative nuclear
$m \rightarrow g$ transition $\Gamma_N$:
\begin{eqnarray}
\beta_{M1} = \frac{\Gamma_{\rm EB}}{\Gamma_N}
\approx \left( 1 - \frac{\omega_1}{\omega_N} \right)^{\!3} \frac{G_2}{3 (2J_t+1)} .
\label{betaM1}
\end{eqnarray}

We use the method of calculation described in detail in Ref.~\cite{PorFla10ThII}.
We consider Th$^+$ as the atom with three valence
electrons above the closed-shell core [1$s^2$, ... ,6$p^6$]
and employ the CI+MBPT approach combining the
configuration-interaction (CI) method in the valence space
with many-body perturbation theory (MBPT) for core polarization
effects~\cite{DzuFlaKoz96b}.
At the first stage we solved Dirac-Hartree-Fock (DHF)
equations~\cite{BraDeyTup77} in $V^{N-3}$ approximation and then
we determined the $5f$, $6d$, $7p$, $7s$, and
$8s$ orbitals from the frozen-core DHF equations. The virtual
orbitals were determined with the help of a recurrent
procedure~\cite{KozPorFla96}. The one-electron basis set included
1$s$--18$s$, 2$p$--17$p$, 3$d$--16$d$, and 4$f$--15$f$ orbitals on
the CI stage.

We formed the configuration spaces
allowing single, double, and triple excitations from the $6d^2 7s$
 configuration
(for the even states) and from the $5f 7s^2$ configuration (for the odd states)
to the 7$s$--13$s$, 7$p$--12$p$, 6$d$--11$d$, and 5$f$--10$f$ shells.
An inclusion of all possible (up to triple) excitations allows us to take into
account most completely the configuration interaction for all
considered states.
 The energies and the wave functions are determined
from the eigenvalue equation in the model space of the valence electrons
\begin{equation}
H_{\mathrm{eff}}(E_{p})\,|\Phi _{p}\rangle =E_{p}\,|\Phi _{p}\rangle,
\label{Eqn_Sh}
\end{equation}
where the effective Hamiltonian is defined as
\begin{equation}
H_{\mathrm{eff}}(E)=H_{\mathrm{FC}}+\Sigma (E).
\label{Eqn_Heff}
\end{equation}
Here $H_{\mathrm{FC}}$ is the relativistic three-electron Hamiltonian in the
frozen core approximation and $\Sigma(E)$ is the energy-dependent
core-polarization correction.

Together with the effective Hamiltonian $H_{\rm eff}$ we
introduce an effective electric-dipole operator $D_{\rm eff}$ and
an operator $(\mathcal{T}_1)_{\rm eff}$ acting in the model
space of valence electrons. These operators were obtained within
the  relativistic random-phase approximation
(RPA)~\cite{DzuKozPor98,KolJohSho82}.
On the stage of solving the RPA equations and calculating diagrams for effective Hamiltonian and
effective operators $D$ and $\mathcal{T}_1$ we used a more extended basis set. It consisted
of $1s$--$22s$, $2p$--$22p$, $3d$--$22d$, $4f$--$22f$, and $5g$--$16g$ orbitals.

{\bf Results and discussion.}
In the following, we assume that the value of $\omega_N$ is between $60\,000$ and
$64\,000$ cm$^{-1}$, which corresponds to the range 7.4--7.9 eV and
suppose that the main contribution to $G_2$ (see \eref{G2}) comes from
intermediate states lying in this range.
As noted above, the odd-parity state $(6d7s7p)\, J=5/2$ at 24874 cm$^{-1}$
is assumed as the initial state $t$.
In~\cite{PorFla10ThIV} we discussed that
the largest value of the coefficient $\beta_{\rm M1}$ is expected if
the initial state and the intermediate state $k$ (see Fig. 2) are connected by
an $E1$ transition. Thus, the states $t$
and $k$ should be of opposite parity and we have to consider the transition
$(6d7s7p, J=5/2) \stackrel{E1}{\longrightarrow} k
                 \stackrel{\mathcal{T}_1}{\longrightarrow} (6d^2 7s, \, J=3/2)$.

Using \eref{G2} we obtain
\begin{eqnarray}
G_2 &\approx& \frac{1}{4} \sum_{\gamma_s}
 \frac{R_{s,J_k=3/2}}{(\omega_{si}-\omega_N + i\Gamma_s/2)^2} \nonumber \\
&+& \frac{1}{6} \sum_{\gamma_p}
 \frac{R_{p,J_k=5/2}}{(\omega_{pi}-\omega_N + i\Gamma_p/2)^2} ,
\label{G2n}
\end{eqnarray}
where the quantity $R_{s,J_k}$ is determined as
\begin{eqnarray}
R_{s,J_k} &\equiv&
      |\langle 6d^2 7s, J=3/2 ||\mathcal{T}_1||\gamma_s J_k \rangle \nonumber \\
&\times& \langle \gamma_s J_k||D|| 6d7s7p, J=5/2 \rangle|^2 .
\label{Rn}
\end{eqnarray}
%
As follows from \eref{betaM1} the equation for $\beta_{M1}$ reads as
\begin{eqnarray}
\beta_{M1} \approx
\frac{1}{18} \left( 1 - \frac{\omega_1}{\omega_N} \right)^{\!3} G_2
\approx 0.012 \times G_2 ,
\label{bM1}
\end{eqnarray}
where we took into account that the quantity $\omega_1/\omega_N \simeq 0.4$ if $\omega_N$ is between
60000 and 64000 cm$^{-1}$ .

High-lying energy levels of Th$^+$ in the range from 60000 to 64000 cm$^{-1}$
were determined theoretically in~\cite{PorFla10ThII}.
In Table~\ref{Tab:1} we list the values of the coefficients $R_{s,J_k}$
found for the even-parity states in the frame of the CI+MBPT+RPA approximation.
The good correspondence between experimental and theoretical levels at lower energies makes us confident that the level structure is complete for the considered electron configurations. We expect that there will be more levels from other configurations but that the coefficients $R$ for configurations with multiple electron excitations are smaller than the dominant ones calculated here.

\begin{figure}
\includegraphics[scale=0.6]{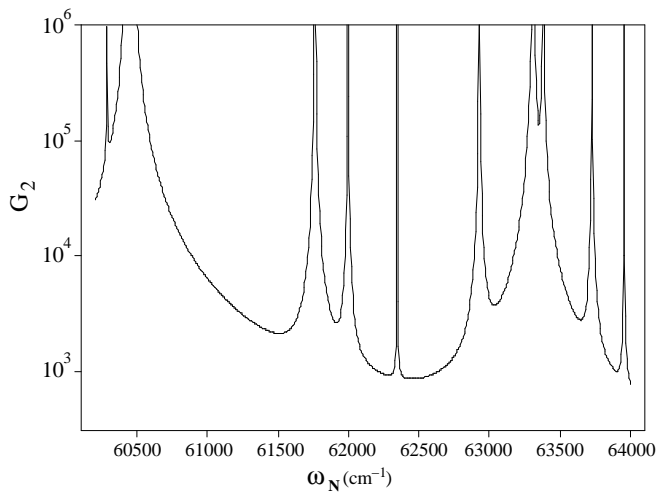}
\caption{Dependence of the coefficient $G_2$ (see \eref{G2n})
on the nuclear transition frequency $\omega_N$ as calculated on the basis of the data listed 
in~\tref{Tab:1}.}
\label{Fig:G2_Th_excit}
\end{figure}

Using \eref{G2n}, the tabulated coefficients $R_{s,J_k}$, and energy values we can
find $G_2$ and $\beta_{M1}$ for a given value of $\omega_N$.
The values of $G_2$ and $\beta_{M1}$ depend critically on the position of $\omega_N$ relative to the electronic levels. This is indicated in
Fig.~\ref{Fig:G2_Th_excit} which shows the variation of $G_2$ with $\omega_N$ in the considered wavenumber range.
As seen from this figure, $G_2 \geq 800$ if $\omega_N$ lies within the range of the electronic energy
levels listed in \tref{Tab:1}. For $\omega_N$ = 7.6 eV~\cite{BecBecBei07}, the $(5f6d7p)\, J=5/2$ state at the calculated value 60462~cm$^{-1}$ yields the main contribution to the EB process and one obtains $G_2 \sim 2700$ and $\beta_{M1} \sim 30$.
When $G_2$ can be calculated on the basis of spectroscopically determined energy levels, it is likely that the dependence of $G_2$ on $\omega_N$ will look somewhat different from Fig.~\ref{Fig:G2_Th_excit}.
Nevertheless, we expect that the EB excitation rate depends on a small number of dominant channels or even only a single one and that
the typical minimum value of the coefficient $G_2$ will remain at the same order of magnitude ($\sim 10^3$).

The excitation probability to the isomeric state due to resonant laser excitation with
$\omega_1 + \omega  = \omega_N$ can be estimated as follows.
We assume that radiation at $\omega$ is produced by a laser source which emits pulses
with 10~mJ energy and with a spectral width of $\Delta \omega = 2\pi \times 3$~GHz at a repetition rate
of 30 Hz. These characteristics can be achieved, for example, with commercially available
frequency-doubled dye lasers as they are used in many laboratory applications.
With focussing to a beam cross section of $0.1 \times 0.1$ mm$^2$, the resulting
time-averaged spectral intensity is
$I \approx 1.5 \times 10^{-3}\,\, {\rm (W/m^2)\,s}$.
Taking into account Eqs.~(\ref{Wa}) and (\ref{betaM1}) and assuming $\beta_{M1} \approx 30$ and the value
$\Gamma_N \simeq  6.6 \times 10^{-4}~{\rm  s}^{-1}$~\cite{PorFla10ThIV},
we find an excitation rate in the range of $W^{\rm in}_{\rm EB} \approx 10\,{\rm  s}^{-1}$.
Significantly larger excitation rates can be expected if a suitable electronic 
transition frequency happens to be very close to $\omega_N$.
\begin{table}
\caption{Calculated energy levels (see text) in the range from 60000 to 64000 cm$^{-1}$
and coefficients $R_{k,J_k}$. $\Delta_k$ is the difference between the energies of the
excited state and the ground state. The notation $y[x]$ means $y \times 10^x$.}

\label{Tab:1}

\begin{ruledtabular}
\begin{tabular}{lccc}
  \multicolumn{1}{l}{$k$}
& \multicolumn{1}{c}{$J_k$}
& \multicolumn{1}{c}{$\Delta_k$\footnotemark[1] (cm$^{-1}$)}
& \multicolumn{1}{c}{$R_{k,J_k} (a.u.)$} \\
\hline
$5f6d7p        $ & 3/2 &  60287  &  3[-4]  \\
$6d7s8s        $ & 5/2 &  60416  &  3[-2]  \\
$5f6d7p        $ & 5/2 &  60462  &  2[-1]  \\
$6d7s8s        $ & 3/2 &  61763  &  2[-3]  \\
$5f^2 6d       $ & 5/2 &  61996  &  5[-4]  \\
$6d^2 8s       $ & 5/2 &  62345  &  1[-5]  \\
$6d7s8s        $ & 3/2 &  62927  &  9[-4]  \\
$6d^2 7d       $ & 5/2 &  63308  &  2[-2]  \\
$6d^2 7d       $ & 3/2 &  63381  &  4[-3]  \\
$6d^2 7d       $ & 3/2 &  63729  &  3[-4]  \\
$6d^2 8s+6d7s8s$ & 5/2 &  63955  &  3[-5]  \\
\end{tabular}
\end{ruledtabular}
\footnotemark[1]{Reference~\cite{PorFla10ThII}}. \\
\end{table}

{\bf Conclusion.} We have suggested a new experimental scheme to excite
the nuclear $^{229g}$Th -- $^{229m}$Th transition in Th$^+$ ions and to
accurately determine its frequency. The scheme relies on an electronic bridge
process that is driven by two incident laser photons whose sum frequency is
resonant with the nuclear transition frequency. Using our previous calculations of
the electronic energy level structure of Th$^+$ in the experimentally
relevant energy range~\cite{PorFla10ThII}, we have estimated the probability of the
investigated two-photon electron bridge process. Assuming that the nuclear transition
energy is close to 7.6 eV, we find that a nuclear excitation probability in the range
of $10 \, {\rm s}^{-1}$ can be obtained with only moderate laser power and
bandwidth requirements.

This work was supported by Australian Research Council.
The work of S.G.P. was supported in part by the Russian Foundation for Basic
Research under Grant No. 08-02-00460-a. The work of E.P. and C.T. was supported
in part by DFG through QUEST.


\begin{thebibliography}{17}
\expandafter\ifx\csname natexlab\endcsname\relax\def\natexlab#1{#1}\fi
\expandafter\ifx\csname bibnamefont\endcsname\relax
  \def\bibnamefont#1{#1}\fi
\expandafter\ifx\csname bibfnamefont\endcsname\relax
  \def\bibfnamefont#1{#1}\fi
\expandafter\ifx\csname citenamefont\endcsname\relax
  \def\citenamefont#1{#1}\fi
\expandafter\ifx\csname url\endcsname\relax
  \def\url#1{\texttt{#1}}\fi
\expandafter\ifx\csname urlprefix\endcsname\relax\def\urlprefix{URL }\fi
\providecommand{\bibinfo}[2]{#2}
\providecommand{\eprint}[2][]{\url{#2}}

\bibitem[{\citenamefont{Peik and Tamm}(2003)}]{PeiTam03}
\bibinfo{author}{\bibfnamefont{E.}~\bibnamefont{Peik}} \bibnamefont{and}
  \bibinfo{author}{\bibfnamefont{C.}~\bibnamefont{Tamm}},
  \bibinfo{journal}{Europhys. Lett.} \textbf{\bibinfo{volume}{61}},
  \bibinfo{pages}{181} (\bibinfo{year}{2003}).

\bibitem[{\citenamefont{Flambaum}(2006)}]{Fla06}
\bibinfo{author}{\bibfnamefont{V.~V.} \bibnamefont{Flambaum}},
  \bibinfo{journal}{Phys. Rev. Lett.} \textbf{\bibinfo{volume}{97}},
  \bibinfo{pages}{092502} (\bibinfo{year}{2006}).

\bibitem[{\citenamefont{Kroger and Reich}(1976)}]{KroRei76}
\bibinfo{author}{\bibfnamefont{L.~A.} \bibnamefont{Kroger}} \bibnamefont{and}
  \bibinfo{author}{\bibfnamefont{C.~W.} \bibnamefont{Reich}},
  \bibinfo{journal}{Nucl. Phys. A} \textbf{\bibinfo{volume}{259}},
  \bibinfo{pages}{29} (\bibinfo{year}{1976}).

\bibitem[{\citenamefont{Reich and Helmer}(1990)}]{ReiHel90}
\bibinfo{author}{\bibfnamefont{C.~W.} \bibnamefont{Reich}} \bibnamefont{and}
  \bibinfo{author}{\bibfnamefont{R.~G.} \bibnamefont{Helmer}},
  \bibinfo{journal}{Phys. Rev. Lett.} \textbf{\bibinfo{volume}{64}},
  \bibinfo{pages}{271} (\bibinfo{year}{1990}).

\bibitem[{\citenamefont{Beck et~al.}(2007)\citenamefont{Beck, Becker,
  Beiersdorfer, Brown, Moody, Wilhelmy, Porter, Kilbourne, and
  Kelley}}]{BecBecBei07}
\bibinfo{author}{\bibfnamefont{B.~R.} \bibnamefont{Beck}},
  \bibinfo{author}{\bibfnamefont{J.~A.} \bibnamefont{Becker}},
  \bibinfo{author}{\bibfnamefont{P.}~\bibnamefont{Beiersdorfer}},
  \bibinfo{author}{\bibfnamefont{G.~V.} \bibnamefont{Brown}},
  \bibinfo{author}{\bibfnamefont{K.~J.} \bibnamefont{Moody}},
  \bibinfo{author}{\bibfnamefont{J.~B.} \bibnamefont{Wilhelmy}},
  \bibinfo{author}{\bibfnamefont{F.~S.} \bibnamefont{Porter}},
  \bibinfo{author}{\bibfnamefont{C.~A.} \bibnamefont{Kilbourne}},
  \bibnamefont{and} \bibinfo{author}{\bibfnamefont{R.~L.}
  \bibnamefont{Kelley}}, \bibinfo{journal}{Phys. Rev. Lett.}
  \textbf{\bibinfo{volume}{98}}, \bibinfo{pages}{142501}
  (\bibinfo{year}{2007}).

\bibitem[{\citenamefont{Tkalya}(2003)}]{Tka03}
\bibinfo{author}{\bibfnamefont{E.~V.} \bibnamefont{Tkalya}},
  \bibinfo{journal}{Phys. Uspekhi} \textbf{\bibinfo{volume}{46}},
  \bibinfo{pages}{315} (\bibinfo{year}{2003}).


\bibitem[{\citenamefont{Porsev and
  Flambaum}(2010{\natexlab{a}})}]{PorFla10ThII}
\bibinfo{author}{\bibfnamefont{S.~G.} \bibnamefont{Porsev}} \bibnamefont{and}
  \bibinfo{author}{\bibfnamefont{V.~V.} \bibnamefont{Flambaum}},
  \bibinfo{journal}{Phys. Rev. A} \textbf{\bibinfo{volume}{81}},
  \bibinfo{pages}{042516} (\bibinfo{year}{2010}{\natexlab{a}}).

\bibitem[{lac()}]{lac-database}
\urlprefix\url{http://www.lac.u-psud.fr/Database/Tab-energy/Thorium/Th-el-dir.%
html}.

\bibitem[{Zim()}]{Zimmermann}
\bibinfo{note}{K. Zimmermann {\it et al.} (to be published)}.

\bibitem[{\citenamefont{Porsev and
  Flambaum}(2010{\natexlab{b}})}]{PorFla10ThIV}
\bibinfo{author}{\bibfnamefont{S.~G.} \bibnamefont{Porsev}} \bibnamefont{and}
  \bibinfo{author}{\bibfnamefont{V.~V.} \bibnamefont{Flambaum}},
  \bibinfo{journal}{Phys. Rev. A} \textbf{\bibinfo{volume}{81}},
  \bibinfo{pages}{032504} (\bibinfo{year}{2010}{\natexlab{b}}).

\bibitem[{\citenamefont{Sobelman}(1979)}]{Sob79}
\bibinfo{author}{\bibfnamefont{I.~I.} \bibnamefont{Sobelman}},
  \emph{\bibinfo{title}{Atomic Spectra And Radiative Transitions}}
  (\bibinfo{publisher}{Springer-Verlag}, \bibinfo{address}{Berlin, Heidelberg,
  New York}, \bibinfo{year}{1979}).

\bibitem[{\citenamefont{Dzuba et~al.}(1996)\citenamefont{Dzuba, Flambaum, and
  Kozlov}}]{DzuFlaKoz96b}
\bibinfo{author}{\bibfnamefont{V.~A.} \bibnamefont{Dzuba}},
  \bibinfo{author}{\bibfnamefont{V.~V.} \bibnamefont{Flambaum}},
  \bibnamefont{and} \bibinfo{author}{\bibfnamefont{M.~G.}
  \bibnamefont{Kozlov}}, \bibinfo{journal}{Phys.\ Rev.\ A}
  \textbf{\bibinfo{volume}{54}}, \bibinfo{pages}{3948} (\bibinfo{year}{1996}).

\bibitem[{\citenamefont{Brattsev et~al.}(1977)\citenamefont{Brattsev, Deyneka,
  and Tupitsyn}}]{BraDeyTup77}
\bibinfo{author}{\bibfnamefont{V.~F.} \bibnamefont{Brattsev}},
  \bibinfo{author}{\bibfnamefont{G.~B.} \bibnamefont{Deyneka}},
  \bibnamefont{and} \bibinfo{author}{\bibfnamefont{I.~I.}
  \bibnamefont{Tupitsyn}}, \bibinfo{journal}{Bull. Acad. Sci. USSR, Phys. Ser.
  (Engl. Transl.)} \textbf{\bibinfo{volume}{41}}, \bibinfo{pages}{173}
  (\bibinfo{year}{1977}).

\bibitem[{\citenamefont{Kozlov et~al.}(1996)\citenamefont{Kozlov, Porsev, and
  Flambaum}}]{KozPorFla96}
\bibinfo{author}{\bibfnamefont{M.~G.} \bibnamefont{Kozlov}},
  \bibinfo{author}{\bibfnamefont{S.~G.} \bibnamefont{Porsev}},
  \bibnamefont{and} \bibinfo{author}{\bibfnamefont{V.~V.}
  \bibnamefont{Flambaum}}, \bibinfo{journal}{J. \ Phys. \ B}
  \textbf{\bibinfo{volume}{29}}, \bibinfo{pages}{689} (\bibinfo{year}{1996}).

\bibitem[{\citenamefont{Dzuba et~al.}(1998)\citenamefont{Dzuba, Kozlov, Porsev,
  and Flambaum}}]{DzuKozPor98}
\bibinfo{author}{\bibfnamefont{V.~A.} \bibnamefont{Dzuba}},
  \bibinfo{author}{\bibfnamefont{M.~G.} \bibnamefont{Kozlov}},
  \bibinfo{author}{\bibfnamefont{S.~G.} \bibnamefont{Porsev}},
  \bibnamefont{and} \bibinfo{author}{\bibfnamefont{V.~V.}
  \bibnamefont{Flambaum}}, \bibinfo{journal}{Zh. \ Eksp. \ Teor. \ Fiz.}
  \textbf{\bibinfo{volume}{114}}, \bibinfo{pages}{1636} (\bibinfo{year}{1998}),
  \bibinfo{note}{[Sov. \ Phys.--JETP {\bf 87} 885, (1998)]}.

\bibitem[{\citenamefont{Kolb et~al.}(1982)\citenamefont{Kolb, Johnson, and
  Shorer}}]{KolJohSho82}
\bibinfo{author}{\bibfnamefont{D.}~\bibnamefont{Kolb}},
  \bibinfo{author}{\bibfnamefont{W.~R.} \bibnamefont{Johnson}},
  \bibnamefont{and} \bibinfo{author}{\bibfnamefont{P.}~\bibnamefont{Shorer}},
  \bibinfo{journal}{Phys. Rev. A} \textbf{\bibinfo{volume}{26}},
  \bibinfo{pages}{19} (\bibinfo{year}{1982}).

\end{thebibliography}

\end{document}